\documentclass[a4paper,pra,twocolumn,showpacs,superscriptaddress,groupedaddress]{revtex4}
\usepackage{ae}
\usepackage{physics}
\usepackage[T1]{fontenc}
\usepackage[ansinew]{inputenc}
\usepackage{amsmath}
\usepackage{amssymb}
\usepackage[caption=false]{subfig}
\usepackage{multirow}
\usepackage{array}
\usepackage{natbib}
\usepackage[]{graphicx}
\usepackage{wrapfig}
\usepackage{makecell}
\usepackage{mathtools}
\usepackage{color}
\usepackage[colorlinks]{hyperref}
\usepackage{lscape}
\newtheorem{theorem}{Theorem}
\graphicspath{ {./image/} }

\DeclarePairedDelimiter\floor{\lfloor}{\rfloor}
\DeclarePairedDelimiter\ceil{\lceil}{\rceil}

\newtheorem{definition}{Definition}
\begin{document}
\title{Catalytic Transformations in Coherence Theory}
\author{Priyabrata Char}
\email{mathpriyabrata@gmail.com}
\affiliation{Department of Applied Mathematics, University of Calcutta, 92 A.P.C Road, Kolkata, 700009, West Bengal, India}
\author{Dipayan Chakraborty}
\email{dipayan.tamluk@gmail.com}
\affiliation{Department of Mathematics, Sukumar Sengupta Mahavidyalaya, Keshpur, Paschim Medinipur, 721150, West Bengal, India}
\author{Amit Bhar}
\email{bharamit79@gmail.com}
\affiliation{Department of Mathematics, Jogesh Chandra Chaudhuri College, 30, Prince Anwar Shah Road, Kolkata 700033, India}
\author{Indrani Chattopadhyay}
\email{icappmath@caluniv.ac.in}
\affiliation{Department of Applied Mathematics, University of Calcutta, 92 A.P.C Road, Kolkata, 700009, West Bengal, India}
\author{Debasis Sarkar}
\email{dsarkar1x@gmail.com, dsappmath@caluniv.ac.in}
\affiliation{Department of Applied Mathematics, University of Calcutta, 92 A.P.C Road, Kolkata, 700009, West Bengal, India}

\begin{abstract}
    In 2016, A. Winter et al.(Physical Review Letters 116 (12) (2016) 120404) provided an operational meaning to relative entropy of coherence and coherence of formation by introducing coherence distillation and dilation protocol in asymptotic setup. Though relative entropy of coherence introduced in 2014 by T. Baumgratz ( Physical Review Letters 113 (14) (2014) 140401) as a coherence measure but it's operational  meaning in single copy setup was unknown so far. Here we have provided relative entropy of coherence (via IO (Incoherent Operations)) and coherence of formation (via IO) and quantum incoherent relative entropy (via LQICC(Local Quantum Incoherent Operations with Classical Communications)) a clear operational significance in single copy setup using the concept of catalyst. We have proved an existential correspondence between asymptotic and catalytic state transformation using IO, LICC(Local Incoherent Operations with Classical Communications) and LQICC. We have also discussed two very important protocols, assisted distillation and quantum incoherent state merging, in single copy setup using catalyst. Monotone property of relative entropy of coherence, coherence of formation and quantum incoherent relative entropy under the catalytic transformation are also discussed here.
\end{abstract}
\date{\today}
\maketitle

\section{Introduction}
Emerging from superposition an important non classical feature of quantum mechanics, the concept of quantum coherence \cite{1} was originated as a fundamental trait in quantum information theory. The resource theory of coherence \cite{32} has come into view by its use in variety of tasks including quantum thermodynamics \cite{2,3,4,5,6}, quantum algorithm, metrology, quantum biology \cite{7,8,9}, spin model \cite{10,11,12}, etc. One of the most useful resource theory is quantum entanglement with which the subject of quantum information theory evolved in the end of last century. Although a wide number of similarities can be found between coherence and entanglement, these two resource theories have many dissimilarities with various features. Absence of unique free operations is one such important difference. Another fundamental difference includes its application for single party system. We now discuss on some basic constructions of this resource theory of coherence. Due to it's basis dependent property, we first fix a reference basis, say, $\{\ket{i}\bra{i}\}_{i=0}^{d-1}$ for a $d$ dimensional system. The set of free states in resource theory of coherence is incoherent states, which are diagonal in reference basis having the form $\rho=\sum_{i} p_i\ket{i}\bra{i}$. The set of operations which do not generate coherence from free states are called free operations. Based on imposing different operational restrictions from different physical motivations, free operations are of different types. Some significant set of free operations are Incoherent Operations (IO), Strictly Incoherent Operations (SIO), Physical Incoherent Operations (PIO). 

Beyond single party the framework of quantum coherence has been extended to the composite quantum systems also. One of the fundamental concept in this framework is the notion of Local Quantum Incoherent Operations with Classical Communications (LQICC) \cite{13,14}. The concept is very similar to the class Local Operations with Classical Communications (LOCC). LQICC differs from LOCC by restricting one party to perform locally incoherent operations only. Further development was done in \cite{15} by including notions of LICC, SI, SQI. The class of operation LICC restricts Alice and Bob to perform locally Incoherent Operations only. 

The concept of catalysis was first introduced in entanglement resource theory \cite{16}. Sometime an impossible transformation using deterministic LOCC $\ket{\psi}\to \ket{\phi}$ can be made possible by using an auxiliary state $\ket{\eta}$ and remarkably the auxiliary state remains unchanged after the transformation. Inspired from chemistry this auxiliary state is called a catalytic state (or catalyst). The concept of catalysis was also used in enhancing success probability in some quantum information processing tasks like entanglement concentration\cite{17,18}. Recently it was shown that catalyst can improve quantum teleportation protocol \cite{19}. By sharing a catalyst state between two parties during quantum teleportation results in a higher fidelity rate. In another ground breaking work \cite{20} T.V.Kondra et al. proved that with the help of entangled catalyst, entropy of entanglement \cite{21,22,23} completely characterizes the quantum state transformation. They have provided a physical meaning of entropy of entanglement in single copy setup using catalysis. Quantum catalysis was also used in other area like quantum thermodynamics \cite{24,25,26,27}, coherence resource theory \cite{28}, purity \cite{6}, asymmetry \cite{30}.    

In this work we have analyzed usefulness of catalysis in some asymptomatic protocols which uses coherence as a resource. Relative entropy of a coherence and Quantum incoherent relative entropy of coherence are two coherence measures related to asymptotic coherence distillation protocols and assisted coherence distillation protocols respectively. There operational meaning for single copy setup was remained open which we have showed also in our work. Section II contains necessary preliminaries and detailed analysis for discussing catalytic state transformation, whereas, Section III contains catalytic assisted distillation and Section IV contains catalytic incoherent quantum state merging respectively. Section V ends with the conclusion.

\section{State Transformation}
For better understanding and visualization of  different features of quantum resource theories, state transformation plays one of the key roles significantly. Several investigations and studies have already been done on state transformation for the finding of vivid pictures of the structure  of quantum states as well as the proper classification of different operations for the resource theories like quantum entanglement and quantum coherence. In this respect, the idea of state transformation of coherence resource theory has been focused in the light of catalytic LICC or LQICC transformations on a composite quantum system and a catalytic IO on a single party system.

Consider a single party or a multi party system $S$ on which a catalytic IO (for single party) or LICC or LQICC (for multi party cases) transformation is defined as $\rho^S\rightarrow \lim\limits_{n\to\infty} Tr_C [\Lambda_n(\rho^S\otimes \tau_{n}^{C})] $ where $C$ is the system of the catalyst and $\tau_{n}^{C}$ is a sequence of catalytic states, $\{\Lambda_n\}$ is a sequence of IO or LICC or LQICC protocols respectively. Further $Tr_S [\Lambda_n(\rho^S\otimes \tau_{n}^{C})]=\tau_{n}^{C} $ should be well satisfied. Essentially the catalyst must decouple from the system after end of the IO or LICC or LQICC processes respectively, for sufficiently large $n$. Mathematically, it is represented as $\lim\limits_{n\to\infty}||\mu_{n}^{SC}-\sigma^{S}\otimes\tau_{n}^{C}||_1=0$ where $\mu_{n}^{SC}=\Lambda_n(\rho^S\otimes \tau_{n}^{C})$ and $\sigma^S$ is the required state.\\

We now start with the development of our concept on catalytic state transformation via LICC or LQICC by recalling an important consequence of asymptotic state transformation under LOCC. 
If a multipartite state $\rho^S $ can be transformed into $\sigma^S $ via asymptotic LOCC with unit rate then there exists a catalytic LOCC protocol transforming $\rho^S $ to $\sigma^S $ \cite{20}. 
This idea truly connects the asymptotic LOCC with unit rate to the existence of catalytic LOCC protocol for the transformation of two multipartite states. In coherence resource theory we are able to find such existence of catalytic LICC or LQICC protocol transformation between two multipartite states when these two states are transformed through asymptotic LICC or LQICC with unit rate. The following theorem is enough to support our claim.
\begin{theorem}
If a multipartite state of $m$ parties $\rho^S $ can be transformed into $\sigma^S $ via asymptotic LICC or LQICC with unit rate then there exists a catalytic LICC or LQICC protocol transforming $\rho^S $ to $\sigma^S $ where $S=A_1A_2...A_m$ . 
\end{theorem}
\textbf{Proof:} Let us assume that each party $A_1,A_2,...,A_m$ of the system $S$ can only do incoherent operations locally, i.e., the system is restricted to LICC operations. The state $\rho^S $ shared between all parties can be transformed to $\sigma^S$ via asymptotic LICC with unit rate. So corresponding to arbitrarily chosen $\epsilon>0$, $\exists \textrm{ } n\in \mathbb{N}$ and a LICC transformation $\Lambda$ such that $\Lambda[(\rho^{S})^{\otimes n}]=\Gamma$ where $D(\Gamma,(\sigma^S)^{\otimes n})<\epsilon$, $D$ is the trace one norm. Let us consider the catalytic state $$\tau=\frac{1}{n}\sum_{k=1}^{n}\rho^{\otimes k-1}\otimes\Gamma_{n-k}\otimes\ket{k}\bra{k}$$
The Hilbert space of the catalyst is $S^{\otimes n-1}\otimes K$ where $K$ is an auxiliary system of dimension n. Let us label the initial system $S$ as $S_1$ and the $n-1$ copies of the same system as $S_2,S_3,...,S_n$. The state of the catalyst acts on $S_2\otimes S_3\otimes...\otimes S_n\otimes K$ and the output state $\Gamma$ acts on $S_1\otimes S_2\otimes...\otimes S_n$; $\Gamma_i=Tr_{A_1A_2...A_{n-i}}\Gamma$ and $\Gamma_0=I$. Now the combined initial state of system and catalyst is 
$$\rho\otimes\tau=\frac{1}{n}\sum_{k=1}^{n}\rho^{\otimes k}\otimes\Gamma_{n-k}\otimes\ket{k}\bra{k}$$
Consider the following LICC protocol.

(i) A rank 1 projective measurement on the auxiliary system $K$ is performed by $A_1$ in the basis $\ket{k}$ and announces the outcome of the measurement to other parties. If $A_1$ obtain the result $n,$ then all other parties will apply the LICC operation $\Lambda$ on their parts. For any other outcome of the measurement they do nothing and the state $\rho\otimes\tau$ transfers to 
$$ \mu_1=\frac{1}{n}\sum_{k=1}^{n-1}\rho^{\otimes k}\otimes\Gamma_{n-k}\otimes\ket{k}\bra{k}+\frac{1}{n}\Gamma\otimes\ket{n}\bra{n} .$$

(ii) $A_1$ applies an incoherent unitary on the auxiliary system $K$ of $ \mu_{1}$ to transfer $\ket{n}\to\ket{1}$ and $\ket{i}\to\ket{i+1}, ~ i= 1, 2, \ldots, n-1 $. Then $\mu_1$ transforms to 
$$\mu_2=\frac{1}{n}\sum_{k=1}^{n}\rho^{\otimes k-1}\otimes\Gamma_{n+1-k}\otimes\ket{k}\bra{k} $$
Note that tracing out the system $S_n$ gives the catalytic state $\tau$.

(iii) All parties apply a SWAP unitary on their parts to transfer $S_n\to S_1$ and $S_i\to S_{i+1}$. Then $\mu_2$ transfers to the final output state $\mu^{SC}$ such that 
\begin{equation}
    D(\mu^{SC},\sigma^S\otimes \tau^C)<2\epsilon \label{eq: distance}
\end{equation}
The proof of inequality \eqref{eq: distance} is same as in the reference \cite{20}.
Note that the operations performed by the party $A_1$ in the above protocol (i)-(iii) can be described as a quantum operation. Hence it is clear that if a asymptotic LQICC $\Lambda$ ( $A_2,A_3,...,A_n$ are restricted to perform incoherent operations only ) exists then there also exists a corresponding catalytic LQICC protocol which transfers $\rho^S$ to $\sigma^S$.   

Now in particular if the system $S$ consists of a single party only, say $A_1$, then the classical communication needed in step (i) of the above protocol will not be necessary and as we are restricted to one party which uses incoherent operation only we have the following theorem.

\begin{theorem}
If a single party state $\rho^A$ can be transformed into $\sigma^A$ via asymptotic Incoherent Operations (IO) with unit rate then there exists a catalytic IO transforming $\rho^A$ to $\sigma^A$.
\end{theorem}
These two results establish the connection between asymptotic LICC or LQICC or IO with unit rate to the existence of catalytic LICC or LQICC or IO state transformation. Our next goal is to check whether relative entropy of coherence is a monotone under catalytic IO.

Relative entropy of coherence \cite{1} is a measure of coherence and it is defined as 
$$C_r(\rho)=\min_{\sigma\in I}S(\rho||\sigma)$$
where $I$ is the set of all incoherent states and $S(\rho||\sigma)=Tr(\rho \log\rho-\rho \log\sigma)$. Again $C_r(\rho)$ can be presented compactly in the following way, i.e., $C_r(\rho)=S(\Delta(\rho ))-S(\rho)$ where $\Delta(\rho)=\sum_i \bra{i}\rho\ket{i}\ket{i}\bra{i}$ and $S$ is the Von Neumann entropy. The relative entropy of coherence is a coherence monotone \cite{1}, convex \cite{1}, additive and super additive \cite{31} and further it is also asymptotically continuous \cite{32}. 
Now our aim of search would be properly described through the following theorem.
\begin{theorem}
If a single party state $\rho^A$ can be transformed into the state $\sigma^A$ via catalytic IO then $C_r(\rho^A)\geq C_r(\sigma^A)$.
\end{theorem}
\textbf{Proof:} We consider an arbitrarily chosen positive number $\epsilon$. Now corresponding to this $\epsilon$ we should find a catalytic state $\tau^{A'}$ and an IO $\Lambda$ such that $\Lambda(\rho^A\otimes\tau^{A'})=\chi^{AA'}$ where $||tr_{A'} (\chi^{AA'})-\sigma^A||_1<\epsilon$ and $\tr_A(\chi_{AA'})=\tau^{A'}$. Now relative entropy of coherence can not increase under IO due to the property of coherence measure. Hence,
\begin{equation}
  C_r(\chi^{AA'})\leq C_r(\rho^A)+C_r(\tau^{A'}) \label{eq:1} 
\end{equation}
Again due to super-additivity of $C_r(\rho)$ we have the following relation.
\begin{equation}
    C_r(tr_{A'} (\chi^{AA'}))+C_r(\tau^{A'})\leq  C_r(\chi^{AA'}) \label{eq:2} 
\end{equation}
The above relations \eqref{eq:1} and \eqref{eq:2} together imply 
\begin{equation}
C_r(tr_{A'} (\chi^{AA'}))\leq C_r(\rho^A) \label{eq:3}
\end{equation}

As relative entropy of coherence is continuous and $||tr_{A'} (\chi^{AA'})-\sigma^A||_1<\epsilon$ therefore from \eqref{eq:3} we can write
$$
  C_r(\sigma^A)\leq C_r(\rho^A)  
$$

In \cite{32} authors proved the following  result regarding asymptotic state transformation under IO.
For any two pure state $\ket{\psi}$ and $\ket{\phi}$ and a rate $R\geq 0$ the asymptotic state transformation via IO is possible if $R<\frac{S(\Delta(\psi))}{S(\Delta(\phi))}$ and impossible if $R>\frac{S(\Delta(\psi))}{S(\Delta(\phi))}$ where $\psi=\ket{\psi}\bra{\psi}$ and $\phi=\ket{\phi}\bra{\phi}$. 
The above result plays an important role  to prove the sufficient part of the following theorem.
\begin{theorem}
A necessary and sufficient condition for a single party pure state $\ket{\psi}$ can be converted into $\ket{\phi}$ via catalytic IO is $S(\Delta(\psi))\geq S(\Delta(\phi))$.
\end{theorem}
\textbf{Proof:} Due to the existence of catalytic IO transformation  in theorem 3 clearly establishes the fact that $C_r(\psi)\geq C_r(\phi)$ which again implies $S(\Delta(\psi))\geq S(\Delta(\phi))$.

Conversely let $S(\Delta(\psi))\geq S(\Delta(\phi))$ holds for any two pure states $\ket{\psi}$ and $\ket{\phi}$. Now the result from \cite{32} which we stated above for $R=1$ and theorem 2  for pure states directly establishes the existence of a catalytic IO protocol that converts $\ket{\psi}$ to $\ket{\phi}$.

The study of state transformations via IO takes a complete picture after the consideration of coherence distillation \cite{32} and dilution \cite{32} protocols. These two protocols are regarded as the natural state transformations in the study of coherence resource theory. Generally in the task of asymptotic coherence distillation using incoherent operations we have to transfer a mixed state $\rho$ to a maximally coherent state \cite{1} $\ket{\phi_2}=\frac{1}{\sqrt{2}}(\ket{0}+\ket{1})$ with a rate $R$. Mathematically we present the fact in the following way:
$$(\rho)^{\otimes n}\xrightarrow[IO]{\epsilon}(\ket{\phi_2})^{\otimes nR} .$$ Again we can transform $(\ket{\phi_2})^{\otimes nR}\xrightarrow[IO]{\epsilon}(\ket{\chi})^{\otimes n}$ such that $C_r(\ket{\chi})=R .$
Hence the total distillation process can be treated as  
\begin{equation}
    (\rho)^{\otimes n}\xrightarrow[IO]{\epsilon}(\ket{\chi})^{\otimes n} \label{eq:tran1} .
\end{equation}
 Theorem 2 with relation \eqref{eq:tran1} clearly elucidates the fact that the possibility of a asymptotic coherence distillation directly establishes the existence of the corresponding catalytic distillation procedure.
 
In this whole process the rate $R$ has a significant role and the optimal value of $R$ clearly determines the success of the procedure. This optimal rate is known as as distillable coherence \cite{32} and is denoted by $C_d(\rho)$. It is defined as
$$C_d(\rho)=sup\{R:\lim_{n\to\infty}(\inf_{\Lambda}||\Lambda[\rho^{\otimes n}]-\phi_{2}^{\otimes\floor{Rn}}||_1)=0\}$$ which is equals to $C_r(\rho)$.

Now the catalytic coherence distillation processes will be more acceptable and applicable if it is proved that the optimal rate is also $C_d(\rho)$. From theorem 3 for catalytic distillation, we can say $C_r(\ket{\chi})\leq C_r(\rho)$ and this directly implies $R\leq C_d(\rho)$, which proves that the optimal rate is $C_d(\rho)$.

Finally our search will be concentrated in characterizing the features of catalytic dilution process under IO. The task of asymptotic coherence dilution using IO can be represented as $\ket{\phi_2}^{\otimes nR} \xrightarrow[IO]{\epsilon} (\sigma)^{\otimes n}$. We can re-describe the process as  
\begin{equation}
    \ket{\chi_1}^{\otimes n} \xrightarrow[IO]{\epsilon} (\sigma)^{\otimes n} \label{eq: tran2}
\end{equation}
such that $C_r(\ket{\chi_1})=R$. Again theorem 2 with \eqref{eq: tran2} makes us to realize the fact that corresponding to asymptotic dilution process there always exists a catalytic dilution process. Both processes of dilution can do same task.
The optimal rate of the asymptotic dilution procedure is well known as coherence cost \cite{32} and is denoted by $C_c (\sigma)$. The coherence cost is defined by 
$$C_c (\sigma)= \inf\{R:\lim_{n\to\infty}(\inf_{\Lambda}||\Lambda[\phi_{2}^{\otimes\floor{Rn}}]-\sigma^{\otimes n}||_1)=0\} .$$
The coherence cost is again equal to the coherence of formation $C_f (\sigma)$ \cite{32} . The coherence of formation is defined by $$C_f(\sigma)=\min\sum_i S(\Delta(\ket{\psi_i}\bra{\psi_i}))$$  subject to $\sigma=\sum_i p_i \ket{\psi_i}\bra{\psi_i}$. The coherence of formation is a coherence monotone, convex \cite{32}  and it is also additive and super additive \cite{33}. The coherence of formation is asymptotically continuous \cite{32}. In general, $C_d(\sigma)\leq C_c(\sigma)$ where equality holds for pure states. 
\begin{theorem}
If a single party state $\rho^A$ can be transformed into the state $\sigma^A$ via catalytic IO then $C_f(\rho^A)\geq C_f(\sigma^A)$.
\end{theorem}
\textbf{Proof:} Let for any $\epsilon>0, \exists \textrm{ a catalytic state } \tau^{A'}$ and an IO $\Lambda$ such that $\Lambda(\rho^A\otimes\tau^{A'})=\chi^{AA'}$ where $||tr_{A'} (\chi^{AA'})-\sigma^A||_1<\epsilon$ and $\tr_A(\chi_{AA'})=\tau^{A'}$. As coherence of formation is a coherence monotone it can not increase under IO. Hence
\begin{equation}
C_f(\chi_{AA'})\leq C_f(\rho^A)+C_f(\tau^{A'}) \label{eq:5}   
\end{equation}
Again as $C_f$ is super additive therefore 
\begin{equation}
C_f(tr_{A'} (\chi^{AA'}))+C_f(\tau^{A'})\leq  C_f(\chi_{AA'}) \label{eq:6}  
\end{equation}
From the above two inequalities \eqref{eq:5} and \eqref{eq:6} we have 
\begin{equation}
C_f(tr_{A'} (\chi^{AA'}))\leq C_f(\rho^A)
\end{equation}

Due to continuity property of relative entropy of coherence and $||tr_{A'} (\chi^{AA'})-\sigma^A||_1<\epsilon$,  we can write
\begin{equation}
  C_f(\sigma^A)\leq C_f(\rho^A)  
\end{equation}
As for a pure states, coherence cost is equal to distillable coherence, therefore taking $\rho^A=\ket{\chi_1}\bra{\chi_1}^A$, we have,
$$C_f(\sigma^A)\leq C_r(\ket{\chi_1}\bra{\chi_1}^A)\implies C_c(\sigma)\leq R .$$
Which proves that the optimal rate for catalytic coherence dilution is also $C_c(\sigma)$.

\section{Assisted distillation}
In this section we pay particular attention to the catalytic assisted distillation via LQICC. Before going through the detailed finding regarding the task we review briefly the standard coherence distillation protocol in the following paragraph.

In coherence assisted distillation \cite{34} tasks, active involvement of at least two parties plays a significant role. Here Alice(A) and Bob(B) can share many copies of a state $\rho^{AB}$ with the objective to obtain coherence as large as possible at Bob's subsystem via a LQICC operation. In this LQICC protocol Bob has the limitation of performing only incoherent operations on his subsystem B. Again the success of such protocols has been assessed through the optimal rate of distillation which is designated as distillable coherence of collaboration \cite{34}. It is defined as $$C_{d}^{A|B}(\rho)=\sup\{R:\lim\limits_{n\to\infty}(\inf||tr_A\Lambda[\rho^{\otimes n}]-\phi_2^{\otimes\floor{Rn}}||_1=0)\}$$ Where infimum is taken over all LQICC operation $\Lambda$. Notice that if $\rho ^{AB}= \rho^{A} \otimes \rho^{B}$, then $C_{d}^{A|B} (\rho)$ reduced to $C_d(\rho^B)$.
 A necessary and sufficient condition for $C_{d}^{A|B}(\rho^{AB})>0$ is the state $\rho^{AB}$ is not quantum incoherent. A quantum incoherent state have the following form $\rho_{AB}=\sum_{i} p_i \sigma^{A}_{i} \otimes \ket{i} \bra{i}^{B}$ where $\sigma_{i}^{A}$ is arbitrary quantum state and $\{\ket{i}^B\}$ is local incoherent basis of Bob's system.
 
We now define quantum incoherent relative entropy \cite{34} of a state $\rho^{AB}$ which is denoted by $C_{r}^{A|B}(\rho^{AB})$ and defined as $$C_{r}^{A|B}(\rho^{AB})=\min\limits_{\sigma^{AB}\in QI}S(\rho^{AB}||\sigma^{AB})=S(\Delta^{B} (\rho^{AB}))-S(\rho^{AB})$$
where $S(\rho)$ is Von Neumann entropy and $S(\rho || \sigma)$ is relative entropy and $\Delta^{B} (\rho^{AB})= \sum (I \otimes \ket{i} \bra{i}) \rho^{AB} (I \otimes \ket{i} \bra{i}) $. $C_{r}^{A|B} (\rho^{AB})$ is monotone under LQICC operation \cite{34}. Quantum incoherent relative entropy is super additive for a four party state $\rho^{ABCD}$ where $B$ and $D$ are restricted to IO operations. The following inequality describe the super additive phenomena.
\begin{equation}
    C_{r}^{A|B}(\rho^{AB})+C_{r}^{C|D}(\rho^{CD})\leq C_{r}^{AC|BD}(\rho^{ABCD})
\end{equation}
For a mixed state $C_{d}^{A|B} (\rho^{AB}) \leq C_{r}^{A|B} (\rho^{AB})$ \cite{34} and for a pure state
\begin{equation}
 C_{d}^{A|B} (\ket{\psi}^{AB})=C_{r}^{A|B} (\ket{\psi}^{AB})=S[\Delta(\psi^{B})]  \label{eq:prop}
\end{equation}
where $\psi^{B}=Tr_{A}(\ket{\psi}\bra{\psi}^{AB})$ \cite{34}.\\

Now consider a pure state $\ket{\psi}^{AB}$ with the purpose that we can create maximally coherent state $\ket{\phi_{2}}=\frac{1}{\sqrt{2}}(\ket{0}+\ket{1})$ in Bob's side with a rate R by applying a asymptotic LQICC operation $\Lambda$. We can readdress the problem as
\begin{equation}
(\ket{\psi}^{AB})^{\otimes n}\xrightarrow[\textrm{LQICC }\Lambda]{\epsilon} (\chi^{AB})^{\otimes n} \label{eq:asstd}
\end{equation}
such that $\tr_{A} (\ket{\chi} \bra{\chi})^{AB}=(\ket{\zeta} \bra{\zeta})^{B}$ with $C_r(\ket{\zeta}\bra{\zeta}^B)=R$. Finally the existence of such LQICC protocol \eqref{eq:asstd} and Theorem 1 for bipartite state directly indicates the existence of a catalytic LQICC protocol through which assisted distillation is possible from $\ket{\psi}^{AB}$.

In the next theorem we will prove that the distillable coherence of collaboration is a monotone under catalytic LQICC operation. 
\begin{theorem}
If a bipartite state $\rho^{AB}$ can be transformed into the state $\sigma^{AB}$ via catalytic LQICC then $C_{r}^{A|B}(\rho^{AB})\geq C_{r}^{A|B}(\sigma^{AB}). $
\end{theorem}
\textbf{Proof:} Let a arbitrary positive number $\epsilon$ is assigned and corresponding to this $\epsilon$ we can find a catalytic state $\tau^{A'B'}$ and a catalytic LQICC $\Lambda$ such that $\nu^{AA'BB'}=\Lambda(\rho^{AB}\otimes\tau^{A'B'})$ where $||tr_{A'B'}(\nu^{AA'BB'})-\sigma^{AB}||_{1}<\epsilon$ and $tr_{AB}(\nu^{AA'BB'})=\tau^{A'B'}$.
As quantum incoherent relative entropy a monotone, it can not increase under LQICC. Hence,
\begin{equation}
C_{r}^{A|B}(\nu^{AA'BB'})\leq C_{r}^{A|B}(\rho^{AB})+C_{r}^{A|B}(\tau^{A'B'})
\end{equation}
Again as quantum incoherent relative entropy is super-additive, we have, 
\begin{equation}
C_{r}^{A|B}(\nu^{AA'BB'})\geq C_{r}^{A|B}(tr_{A'B'}(\nu^{AA'BB'}))+C_{r}^{A|B}(\tau^{A'B'})
\end{equation}
From the above two inequalities we have,
\begin{equation}
C_{r}^{A|B}(tr_{A'B'}(\nu^{AA'BB'}))\leq  C_{r}^{A|B}(\rho^{AB})     
\end{equation}
As quantum incoherent relative entropy is continuous \cite{34} and $||tr_{A'B'}(\nu^{AA'BB'})-\sigma^{AB}||_{1}<\epsilon$, therefore we can write 
\begin{equation}
C_{r}^{A|B}(\rho^{AB})\geq C_{r}^{A|B}(\sigma^{AB})  
\end{equation}

Now we will prove that the optimal rate for catalytic assisted distillation is also $C_{d}^{A|B} (\ket{\psi})\bra{\psi}^{AB}$. If we transfer $\ket{\psi}^{AB}$ to $\ket{\chi}^{AB}$ such that $\tr_{A} (\ket{\chi} \bra{\chi})^{AB}=(\ket{\zeta} \bra{\zeta})^{B}$ with $C_r(\ket{\zeta}\bra{\zeta}^B)=R$ then by theorem 6 we have,
%As Quantum Incoherent  relative entropy is monotone and additive under LQICC operation. Therefore
%$$C_{r}^{A|B} (\ket{\psi}^{AB}\otimes \tau^{AB})\geq C_{r}^{A|B} (\ket{\chi}^{AB}\otimes \tau^{AB})$$
$$C_{r}^{A|B} (\ket{\psi}\bra{\psi}^{AB})\geq C_{r}^{A|B} (\ket{\chi}\bra{\chi}^{AB})$$
$$\implies C_{d}^{A|B}(\ket{\psi}\bra{\psi}^{AB})\geq S(\Delta(\ket{\zeta}\bra{\zeta}^B)) \textrm{ [ using }\eqref{eq:prop}\textrm{ ]}$$
$$\implies C_{d}^{A|B}(\ket{\psi}\bra{\psi}^{AB})\geq C_r(\ket{\zeta}\bra{\zeta}^B)$$
$$\implies C_{d}^{A|B}(\ket{\psi}\bra{\psi}^{AB})\geq R ,$$
Which proves our claim.

Finally our aim of search will be focused on the catalytic incoherent quantum state merging protocol via LQICC.
\section{Incoherent quantum state merging}
Before presenting our work we look at the standard quantum state merging at a glance.

In standard quantum state merging \cite{35,36}, Alice, Bob, Referee, shares asymptotically many copies of $\ket{\psi}^{ABR}$ and their goal is to transfer Alice's part to Bob by applying LOCC between Alice and Bob while correlation with Referee is preserved. So $\ket{\psi}^{ABR}$ and $\ket{\psi}^{BBR}$ is the same state. Now quantum conditional entropy of the state is defined as $S(A|B)_{\psi}=S(\psi^{AB})-S(\psi^B)$, where $\psi^{AB}=Tr_{R}(\ket{\psi}\bra{\psi}^{ABR})$ and $\psi^{B}=Tr_{AR}(\ket{\psi}\bra{\psi}^{ABR})$. When $S(A|B)_{\psi}>0$ quantum state merging is possible if we additionally supply singlets at rate $S(A|B)_{\psi}$. If $S(A|B)_{\psi}\leq 0$ then quantum state merging is possible without any additional resource, in fact when $S(A|B)_{\psi}<0$ then we additionally gain singlets at the rate $-S(A|B)_{\psi}$. \\

In the task of Incoherent Quantum State Merging (IQSM) \cite{37} the resources are quantified by the pair $(E,C)$ where $E$ is entanglement and $C$ is coherence and LQICC is the operation used by Alice and Bob with Bob has access to incoherent operations only. The fascinating fact about IQSM is that we can not gain entanglement and coherence at same time which is guaranteed by the following  inequality \cite{37}
    $$ E+C \geq S(  I^{R}\otimes \Delta^{AB}[\rho^{RAB}]) -S(  I^{RA}\otimes \Delta^{B}[\rho^{RAB}]).$$
 When the IQSM protocol consumes entanglement(coherence) then $E(C)$ is positive and gaining entanglement(or, coherence) in IQSM means E( or, C) is negative. Now, we have the following important definitions.
 
\begin{definition}
Let $E_i \textrm{ and }E_t$ is the initial and final entanglement rate shared between Alice and Bob and similarly $C_i \textrm{ and } C_t$ is initial and final local coherence of Bob. An entanglement-coherence pair $(E,C)$ is achievable if $\exists $ numbers $E_i,E_t,C_i,C_t$ with $E=E_i-E_t$ and $C=C_i-C_t$ such that for any $\epsilon>0$ and any $\delta>0$, $\exists m\in\mathbb{N}$ such that $\forall n\geq m$, $\exists$ a LQICC protocol $\Lambda$ between Alice and Bob such that \\
$
\frac{1}{2}||\Lambda[\rho_{i}^{\otimes n}\otimes \ket{\phi^{+}}\bra{\phi^{+}}^{\otimes \ceil{(E_i+\delta)n}}\otimes \ket{\phi_{2}}\bra{\phi_2}^{\ceil{(C_i+\delta)n}}] \\ -\rho_{t}^{\otimes n}\otimes \ket{\phi^{+}}\bra{\phi^{+}}^{\otimes \floor{(E_tn}}\otimes \ket{\phi_{2}}\bra{\phi_2}^{\floor{C_tn}} ||_1< \epsilon ,
$
where $\rho_i=\rho^{RAB}\otimes\ket{0}\bra{0}^{\Tilde{B}}$ is total initial state, $\Tilde{B}$ is the additional system of Bob with $\dim(\Tilde{B})=\dim(A)$, $\ket{\phi^+}=\frac{1}{\sqrt{2}}(\ket{00}+\ket{11})$, $\ket{\psi_2}=\frac{1}{\sqrt{2}}(\ket{0}+\ket{1})$ and $\rho_t=\rho^{R\Tilde{B}B}\otimes\ket{0}\bra{0}$; $\rho_i$ and $\rho_t$ is the same state up to relabelling of $A$ and $\Tilde{B}.$ \cite{37} 
\end{definition}

\begin{definition}
A pair $(E,C)$ is called optimal if (1) the pair is achievable and (2) $(E^{'},C)$ and $(E,C^{'})$ are not achievable for any $E^{'}\leq E \textrm{ or } C^{'}\leq C$. \cite{37}
\end{definition}
Some optimal pair of asymptotic IQSM protocol are $(E_0,0)$, $(E_{min}, C_{max})$, etc. Now we state the following theorem\cite{37}.
\begin{theorem}
A pure state $\ket{\psi}^{RAB}$ with the optimal pair $(E_0,0)$ can be merged via asymptotic IQSM protocol where $E_0=S(\bar{\rho}^{AB})-S(\bar{\rho}^{B})$ where $\bar{\rho}^{AB}=\Delta(\rho^{AB})$, $\bar{\rho}^{B}=\Delta(\rho^{B})$, $\rho^{AB}=Tr_{R} \ket{\psi} \bra{\psi}^{RAB}$, $\rho^{B}=Tr_{RA} \ket{\psi} \bra{\psi}^{RAB}$ \cite{37}.

\end{theorem} 
Now let us consider a IQSM protocol via asymptotic LQICC with achievable pair $(E_{0},0)$. Let us define $\ket{\psi_{1}}^{RA\Tilde{A}B\tilde{B}}=\ket{\psi}^{RAB} \otimes \ket{\chi}^{\tilde{A}\tilde{B}}$ with $S(Tr_{\tilde{A}} \ket{\chi} \bra{\chi} ^{\tilde{A}\tilde{B}})=E_0$ 
As additional resource have been supplied then $\exists$ an asymptotic LQICC protocol such that 
\begin{equation}
    (\ket{\psi_{1}}^{RA\tilde{A}B\tilde{B}})^{\otimes n}\xrightarrow[LQICC]{\epsilon} (\ket{\psi}^{RB^{'}B})^{\otimes n} \label{eq:IQSM}
\end{equation}
An immediate consequence of theorem 1 and above equation \eqref{eq:IQSM} is that catalytic IQSM is possible when Alice and Bob have shared an additional resource $\ket{\chi}^{\tilde{A}\tilde{B}}$. Next we will prove that $E_0$ is also the minimum entanglement required for catalytic IQSM.

Let us consider a state $\ket{\chi}=\sum_{i}\sqrt{\lambda_{i}}\ket{ii}$ in the Schmidt basis with $S(Tr_{\tilde{A}} \ket{\chi} \bra{\chi} ^{\tilde{A}\tilde{B}})=R$. We fix the basis of the system $AB$ in Schmidt basis. This will also fulfill basis dependency criteria for resource theory of coherence. As we take $\ket{\chi}$ in Schmidt basis then $\Delta^{\tilde{B}}(Tr_{\tilde{A}} (\ket{\chi} \bra{\chi} ^{\tilde{A}\tilde{B}})))=(Tr_{\tilde{A}} \ket{\chi} \bra{\chi} ^{\tilde{A}\tilde{B}})$. Using $\eqref{eq:prop}$, we have $C _{r}^{\tilde{A}|\tilde{B}}(\ket{\chi}\bra{\chi}^{\tilde{A}\tilde{B}})=S(\Delta^{\tilde{B}}(Tr_{\tilde{A}} (\ket{\chi} \bra{\chi} ^{\tilde{A}\tilde{B}}))) $. Therefore, $C _{r}^{\tilde{A}|\tilde{B}}(\ket{\chi}\bra{\chi}^{\tilde{A}\tilde{B}})=R .$
\begin{widetext}
$$C_{r}^{RA\tilde{A}|B\tilde{B}}(\ket{\psi}\bra{\psi}^{RAB}\otimes \ket{\chi} \bra{\chi}^{\tilde{A}\tilde{B}})=C _{r}^{RA|B}(\ket{\psi}\bra{\psi}^{RAB}) + C _{r}^{\tilde{A}|\tilde{B}}(\ket{\chi}\bra{\chi}^{\tilde{A}\tilde{B}}) \textrm{ ( by additive property )}$$ 
$$\textrm{So, } C_{r}^{RA\tilde{A}|B\tilde{B}}(\ket{\psi}\bra{\psi}^{RAB}\otimes \ket{\chi} \bra{\chi}^{\tilde{A}\tilde{B}})=S(\Delta(\rho^B))+C _{r}^{\tilde{A}|\tilde{B}}(\ket{\chi}\bra{\chi}^{\tilde{A}\tilde{B}}) \textrm{ ( using \eqref{eq:prop} )}$$ 
$$\textrm{ Again using, \eqref{eq:prop},  }C _{r}^{R|B}(\ket{\psi}\bra{\psi}^{RB^{'}B})=S(\Delta(\rho^{B^{'}B})).$$
$$\textrm{Now as } \ket{\psi}^{RB^{'}B} \textrm{ is same as } \ket{\psi}^{RAB} \textrm{up to relabelling of }A \textrm{ as } B^{'} \textrm{ then, }S(\Delta(\rho^{B^{'}B}))=S(\Delta(\rho^{AB}))$$.  $$\textrm{ So,  }C _{r}^{R|B}(\ket{\psi}\bra{\psi}^{RB^{'}B})=S(\Delta(\rho^{AB})).$$
$$\textrm{Next by theorem 6, } C_{r}^{RA\tilde{A}|B\tilde{B}}(\ket{\psi}\bra{\psi}^{RAB}\otimes \ket{\chi} \bra{\chi}^{\tilde{A}\tilde{B}})\geq C_{r}^{R|B}(\ket{\psi}\bra{\psi}^{RBB'}).$$
$$ \implies S(\Delta(\rho^B))+C _{r}^{\tilde{A}|\tilde{B}}(\ket{\chi}\bra{\chi}^{\tilde{A}\tilde{B}})\geq S(\Delta(\rho^{AB}))$$
$$ \implies C _{r}^{\tilde{A}|\tilde{B}}(\ket{\chi}^{\tilde{A}\tilde{B}})\geq S(\Delta(\rho^{AB}))-S(\Delta(\rho^{B}))$$
$$ \implies R\geq E_0$$
\end{widetext}

\section{Conclusion}
In summary it is nicely perceived that relative entropy of coherence is one of the most useful measure to quantify the total amount of coherence in a single party. It's operational significance is related to the asymptotic coherence distillation protocol and for single copy setup it's role remain unknown. Similar to relative entropy of coherence, quantum incoherent relative entropy plays a key role in multiparty scenario as it is the optimal rate in the asymptotic assisted distillation protocol using LQICC. In our paper we have succeeded to enlighten their operational significance in single copy by constructing a correspondence between asymptotic and catalytic state conversion via IO or LQICC. We have showed that corresponding to an asymptotic IO or LICC or LQICC there always exists a corresponding catalytic IO or LICC or LQICC. Thereafter we have proved that relative entropy of coherence, coherence of formation and  quantum incoherent relative entropy of coherence is also monotone under catalytic transformation. We have provided a necessary and sufficient condition for pure state transformation under catalytic IO. We have also showed that distillation, dilution, assisted distillation, quantum incoherent state merging is also possible in single copy setup with the help of entangled catalyst as a particular application of state transformation. This process has also opened the door to repeat a transformation process with arbitrary small error using only a single copy input state and a catalyst, which will be recovered after the end of the process. Consequently the technique for using such catalytic protocol have also prevented the loss of resource which indicates significant and efficient directions of any resource theoretical approach. We have also proved that the optimal rates of above described processes in single copy setup with the help of catalyst is same as the asymptotic setup which also highlights the importance  of catalytic protocol.\\
Note: After submitting our work in arxiv we have found a related work in arxiv:2106.12592 (intimated by the authors)\cite{related} where they have described monotone property of certain resource measure and relation between asymptotic and single copy state transformation using help of marginal or correlated catalyst. 

\section*{Acknowledgements}
Priyabrata Char acknowledges the support from Department of Science and Technology(Inspire), New Delhi,
India. The authors Debasis Sarkar and Indrani Chattopadhyay acknowledges it as Quest initiatives.

\end{document}